\documentclass[modern]{aastex61}
\usepackage{mathtools}
\usepackage{epstopdf}
\usepackage{amsmath}
\usepackage{color}
\definecolor{Comm}{rgb}{1,0,0}
\received{July 1, 2017}
\revised{September 27, 2017}
\accepted{June 7, 207}
\published{July 28, 2017  in PASP}

\shorttitle{Optimization of exposure time division}
\shortauthors{Popowicz et al.}

\begin{document}
\title{Optimization of exposure time division for wide field observations}
\correspondingauthor{Adam Popowicz}
\email{apopowicz@polsl.pl}

\author{Adam Popowicz}
\affil{Institute of Automatic Control, Silesian University of Technology, Gliwice, Poland}

\author{Aleksander R. Kurek}
\affiliation{Astronomical Observatory of the Jagiellonian University, Cracow, Poland}

\begin{abstract}
	The optical observations of wide fields of view encounter the problem of selection of best exposure time. As there are usually plenty of objects observed simultaneously, the quality of photometry of the brightest ones is always better than of the dimmer ones. Frequently all of them are equally interesting for the astronomers and thus it is desired to have all of them measured with the highest possible accuracy.
	
	In this paper we present a novel optimization algorithm dedicated for the division of exposure time into sub-exposures, which allows to perform photometry with more balanced noise budget. Thanks to the proposed technique, the photometric precision of dimmer objects is increased at the expense of the measurement fidelity of the brightest ones. We tested the method on real observations using two telescope setups demonstrating its usefulness and good agreement with the theoretical expectations. The main application of our approach is a wide range of sky surveys, including the ones performed by the space telescopes. The method can be applied for planning virtually any photometric observations, in which the objects of interest show a wide range of magnitudes.
\end{abstract}


\section{Introduction}
\label{sec:intro}
Wide field sky surveys are an integral part of astronomical research. There are plenty of them performed so far, as well as new ones are planned\footnote{For a well known list of large surveys, see \href{http://www.astro.ljmu.ac.uk/~ikb/research/galaxy-redshift-surveys.html}{Ivan K. Baldry webpage} or one of its custom, updated versions, e.g. \href{http://enaa2013.oal.ul.pt/wp-content/uploads/sites/2/2013/04/JarleBrinchmann.pdf}{Euclid Legacy Science document}, p.\,27.}. Numerous findings and discoveries would not be possible without this type of observations. To get the full picture, see e.g. \href{http://vipers.inaf.it/}{VIPERS} survey \href{http://vipers.inaf.it/papers.html}{paper list} and analogical, even longer, lists for \href{http://www.sdss3.org/science/publications.php}{SDSS III} or \href{http://www.sdss.org/science/publications/}{SDSS IV} surveys. Even some photographic plates-based surveys, performed tens of years ago, are still in use, e.g. \href{http://archive.stsci.edu/cgi-bin/dss_form}{Digitized Sky Survey} (DSS). Moreover, the amount of data gathered in surveys is continuously increasing, putting ever higher demands on algorithms and data reduction techniques, e.g. the \href{https://www.lsst.org/}{Large Synoptic Survey Telescope} (LSST) will collect 15 TB of data per night \citep{LSSTdata}. This changes drastically the proportion between the budget for instrumentation (telescopes, domes, mounts, cameras etc.) and the cost of computers\footnote{See \citet{largeObs} for an overview of large observatory planning and construction.}. Some of the planned survey projects, like the LSST, indicate that sometimes the telescope itself can be only a part of a gigantic computation facility dedicated for data processing \citep{LSSToverview1, LSSToverview2}.

Somewhere in between the instrumentation and the data processing, the strategy of proper management of observing time takes place. Its importance cannot be underestimated, since it affects directly the quality of retrieved data, which is going to be analyzed, surely, for the next tens of years. The management of time is a complex task, in which plenty of various aspects have to be considered. There are a lot of conflicting interests, like sky coverage, time resolution and photometric quality, for which a trade off has to be achieved \citep{largeObs}. Importantly, the time of space observatories, like the \href{http://hubblesite.org/}{Hubble Space Telescope} \citep{HSTimpact} or \href{http://sci.esa.int/herschel/}{Herschel Space Observatory} \citep{Herschel}, is especially valuable and has to be well optimized to maximize their scientific output.

We concentrate here on one particular question, which emerges while observing objects showing wide range of magnitudes: is it better to (a) perform only shorter exposures to include brightest stars in all images, or (b) to take longer ones while registering only several frames for the brightest sources, or maybe (c) to divide the available time in some specific way? Since it is frequent that all the stars in the field are interesting for astronomers, and due to the fact that the density of objects increases exponentially with decreasing their intensity (so-called inverse-square law), the proper optimization of the exposure times selection seems to have significant influence on the final outcome from the photometric series or from surveys. It is obvious that if short exposures are taken, the dimmer sources will be measured with significantly lower quality than the stronger ones. Also, the impact from readout noise will be higher and more time will be wasted on the detector readout, which can be relatively high for wide field cameras. On the other hand, if long exposures are used, the photometry of dimmer stars will be enhanced and less time will be spent on the readout, while the brightest stars will be oversaturated. A natural solution is to take longer and shorter exposures alternately so that one can measure both objects. But the question arises, how to perform the selection of exposure times if the stars show a range of magnitudes, as it is typical in wide fields of view? This is in fact a very complicated issue and the answer will depend on (1) what magnitudes are currently observed in the field, (2) what is the required time resolution (and/or available observing time for the measurement of a portion of sky) and (3) what is the desirable photometric precision. 

Fortunately, one can determine at least what is the time resolution dictated either by time or sky coverage constraints (surveys) or by the frequencies of flux fluctuations of the observed objects (photometric series). Having this information, i.e. knowing the total exposure time which is provided for obtaining a single photometric point, the problem described above can be reduced to the following question: how to divide the available exposure time, so that the quality of photometry for a wide range of magnitudes is balanced in desired way. In this paper we assume that the available time to obtain a single photometric point is long enough (and/or telescope aperture is big enough) so that many of observed sources saturate the detector. Otherwise, the time would be spent on a single exposure and there would be no sense in time division (this situation may appear if a very small telescope or very narrow photometric filter is employed). We also assume that the observations are made in a wide field and all the observed stars are potentially interesting for the observer (i.e. it is not a single-object photometry). It is in fact a case for sky-surveys, in which a telescope obtains a single photometric measurement (by taking several sub-exposures) and is moved to the next sky patch.

In this paper we show the method for solving this complicated and highly nonlinear problem. Utilizing the genetic approach, we developed the routine which allows for optimal balancing the noise budget \citep[Sec. 3.2.3 therein]{largeObs} among a wide range of magnitudes of observed objects. Thanks to our method, the photometric quality of the dimmer objects is increased at the cost of decreasing the quality of brighter ones, which are in fact frequently limited by multiplicative noise due to e.g imperfect flat fielding \citep{Flatfielding} or scintillation noise. The efficiency of our approach was examined by real on-sky experiment on random stellar fields, in various seeing conditions, and using two telescopes of significantly different aperture and focal length.

The paper is organized as follows. In Section \ref{sec:basics} we show the analysis of noise sources in CCD photometry. Then, in Sec.~\ref{sec:efects} we provide the derived formula for the quality estimation of the measurement composed of several exposures. The next section presents the genetic approach which aims at the optimal division of total exposure time to balance the noise budget for a range of magnitudes. In Sec.~\ref{sec:Experiment} we show the results of on-sky experiments which were obtained with a use of our algorithm and using various astronomical instrumentation . We conclude in Sec. \ref{sec:Conclusions}.
\\
\\
\section{Noise formalism}
\label{sec:basics}
To explain the significance of proper division of observing time, first we derive a fundamental formulas which show the limits of measurement accuracy. We start with the estimation of the noise variance present in the photometric measurements, which is given by:
\begin{equation}
	\sigma^2 = \sigma^2_p + \sigma^2_b + \sigma^2_d + \sigma^2_{RON}  + \sigma^2_s + \sigma^2_{ff},
	\label{noiseTerms}
\end{equation}
where $\sigma^2_p$, $\sigma^2_b$ and $\sigma^2_d$ are the noise terms originating in Poisson distribution of counts from the source, the background and from the dark current, respectively, $\sigma_{RON}^2$ is the impact of readout noise, $\sigma^2_s$ is the scintillation noise and $\sigma^2_{ff}$ is the flat-fielding noise. As all the noise terms are independent, a total variance can be calculated as a simple sum of all variances.

Assuming the utilization of aperture photometry, in which the aperture of $A_{pix}$ pixels cover whole stellar profile, the first three variances can be calculated as follows:
\begin{equation}
	\sigma^2_s=dS~t_e, 
\end{equation}
\begin{equation}
	\sigma^2_b=A_{pix}~dB~t_e + A_{pix}^2(\sigma_{r}^2+dB~t_e)/B_{pix}, 
	\label{Back_eq}
\end{equation}
\begin{equation}
	\sigma^2_d=A_{pix}~ dD~t_e, 
\end{equation}
where $t_e$ is exposure time,  $\sigma_{r}$ is the readout noise of a single pixel expressed in electrons, $dS$ is the number of counts registered from a source per second $[e^- /s]$, $dB$ is the number of counts from the background per pixel and per second $[e^-/ \textrm{pix}/ s]$, $dD$ is the average number of electrons excited by the thermal activity (i.e. the dark current) in a pixel per second $[e^-/ \textrm{pix}/s]$. While the factors $dS$ and $dB$ are dependent on the intensity of currently observed object and on the sky brightness, $dD$ has to be estimated from the calibration dark frames (in strongly cooled cameras, this term is usually negligibly small). 

An additional comment is needed to explain the eq.~\ref{Back_eq}. This term consists of two components which correspond to, respectively, (first) the Poisson noise of the counts from background within aperture $A$ and (second) the inaccuracy of background subtraction. Let us elaborate on the latter. In the aperture photometry the estimation of expected intensity of the background is performed in the outer ring consisting $B_{pix}$ pixels, and such an offset is subsequently subtracted from all $A$ pixels in the aperture. The estimated background level has a variance which includes the readout and the Poisson noise, both divided by $B_{pix}$ pixels due to the averaging of pixels in the outer ring:
\begin{equation}
	\sigma_{BL}^2=(B_{pix}~\sigma_{RON}^2+B_{pix}~dB~t_e)/B_{pix}^2 = (\sigma_{RON}^2+dB~t_e)/B_{pix}. 
\end{equation}
However, since the estimated background level is subtracted from all pixels in the aperture $A$, the standard deviation of the photometric result is multiplied by $A_{pix}$ pixels giving $A_{pix}\,\sigma_{BL}$. After squaring, this eventually leads to the second component of the background-related variation (eq.~\ref{Back_eq}). To minimize this noise factor, one should use as large background ring as it is possible. Obviously, this size is always limited by possible background fluctuations and by the presence of nearby stars.

The readout noise in a single pixel, $\sigma_{r}$, can be easily determined from bias frames or it is provided by the camera vendor. Since in the aperture photometry we sum the pixels over the aperture, the total noise becomes:
\begin{equation}
	\sigma_{RON}^2 = A_{pix}~\sigma_r^2.
\end{equation}
The next term, scintillation noise $\sigma^2_s$, originates in turbulent nature of the atmosphere and has been recently investigated by \cite{Osborn}. The authors derived the following formula for relative variance of scintillation noise:
\begin{equation}
	\sigma^2_{s~\textrm{norm}} = 10^{-5}~C^2_Y~D^{-4/3}~t_e^{-1}~ (\textrm{cos} \gamma)^{-3}~\textrm{exp}(-2h_{\textrm{obs}}/H),
\end{equation}
which can be further multiplied by a signal to achieve the variance expressed in electrons:
\begin{equation}
	\sigma^2_s = (dS~ t_e)^2 ~\sigma^2_{s~\textrm{norm}},
\end{equation}
where: $D$ is telescope diameter, $\gamma$ is the zenith distance, $h_{\textrm{obs}}$ is the altitude of the observatory, $H$ is the scale height of the atmospheric turbulence (generally assumed to be $\sim$8000\,m), $C_Y$ is the scaling coefficient which can be demerited from turbulence profilers (SCIDARs) and was estimated between 1.30$\sim$1.67 for best observing sites (see Tab.~1 in the original work by \cite{Osborn}). As we are going to modify the exposure time, it is more convenient for us to represent $\sigma^2_s$ in the following form:
\begin{equation}
	\begin{split}
		\sigma^2_s &= dS^2~ t_e  ~\sigma^2_s \prime, \\
		\sigma^2_s \prime &= 10^{-5}~C^2_Y~D^{-4/3}~ (\textrm{cos} \gamma)^{-3}~\textrm{exp}(-2h_{\textrm{obs}}/H).
		\label{scinEQ}
	\end{split}
\end{equation}

The scintillation noise has the multiplicative nature, which means that it is linearly dependent on the signal. It is responsible for the appearance of noise floor which limits the photometric accuracy of the brightest objects. The absolute value of this noise can be somehow mitigated by increasing telescope aperture or extending exposure time. Importantly, the scintillation noise is not present in observations made by the space telescopes.

The last noise term, flat-fielding error, is also linearly dependent on the collected charge and can be minimized by utilizing high-quality flat-field calibration frames. If a telescope tracks perfectly and the stellar profiles are identical in all images, then there is no additional noise introduced to the photometric lightcurves. Otherwise, if the profile's position changes between exposures (either because of telescope tracking errors or due to the atmospheric effects), two possible scenarios take place. First one, when the shifting is only slight (say, a few pixels), the parts of stellar profiles are sampled by different pixels. In such a case one should expect that the inaccuracies of the estimation of pixels sensitivities will introduce the following errors:
\begin{equation}
	\sigma^2_{ff1} = (dS~t_e)^2\sum_{(x,y) \in P} I_n(x,y)^2r_{1}^2,
	\label{eqff1}
\end{equation}
where $r_{1}$ is the average uncertainty of the estimation of pixel sensitivity in a local neighborhood, $I_n(x,y)$ is the normalized intensity of stellar profile at pixel $(x,y)$. The normalizations here means that the profile fulfills $\sum_{(x,y) \in P}\, I_n(x,y) = 1$. The formula is a simple weighted average of variances from all pixels included in a stellar profile and it is assumed that the errors of sensitivity estimation in neighboring pixels are uncorrelated. The magnitude of such noise is usually very small and can be neglected except for the profiles which cover very few pixels (poor sampling, complex profiles, usually in extremely wide field of view). However, the prediction with this formula can be tricky, since for longer exposure times, tracking errors are averaged, leading to wider and more stable profiles. 

The second scenario appears if the stars are significantly shifted between exposures, so that they fall on completely different parts of CCD from frame to frame. This is a typical problem of the surveys employing sky scanning, but was also encountered for Hubble data \citep{Flatfielding}. It can become a signifiant issue if one combines observations from several nights. In such a case the errors are usually much greater and the following approximation can be used:
\begin{equation}
	\sigma^2_{ff2} = (dS~t_e)^2~r_{2}^2,
	\label{eqff2}
\end{equation}
where $r_{2}$ is the maximal uncertainty of the estimation of pixel sensitivity over whole image plane, which is related to non-ideal flat fielding. It is to be noted that previously defined $r_{1}$ and $r_{2}$ have significantly different magnitude. While the first corresponds to the flat-fielding uncertainties between the pixels included in the stellar profiles (or in its close proximity), the second estimates the same effect but over whole sensor.

Due to the complex and instrumental-dependent nature of flat-field effects, it is difficult to determine a single noise level globally. Depending on the observing strategy, one of presented formulas, \ref{eqff1} or \ref{eqff2}, should be used. Importantly, both of them allow for estimating the upper bound of the noise rather then predicting the expected value. The flat-field effects, similarly to the scintillation noise, predominate the other noise sources when observing bright stars. Together with scintillation noise (only for ground-based telescopes), they prevent observers from reaching the photon-noise-limited accuracy for the brightest objects.

As an addition to the noise sources mentioned above, there are other types which emerge only in very specific observational techniques. They usually have impulsive nature and have been recently intensively examined and explained by the authors in \citep{ImpulsiceNoise}. The impulsive noise appears e.g. in EMCCD cameras (clock-induced charge, CIC) or in instruments working in high-radiation space environment (cosmic-rays, hot pixels, charge transfer inefficiencies CTI, see e.g. \cite{BRITEPASP}). Unfortunately, it is impossible to quantitatively assess its contribution to the noise budget since this can be either multiplicative (CTI) or additive (hot pixels) type of noise. Due to this fact and also because this noise is limited to very specific applications, we decided not to include it in the main formula~\ref{noiseTerms}.

\begin{figure}
	\centering
	\includegraphics[width=\textwidth]{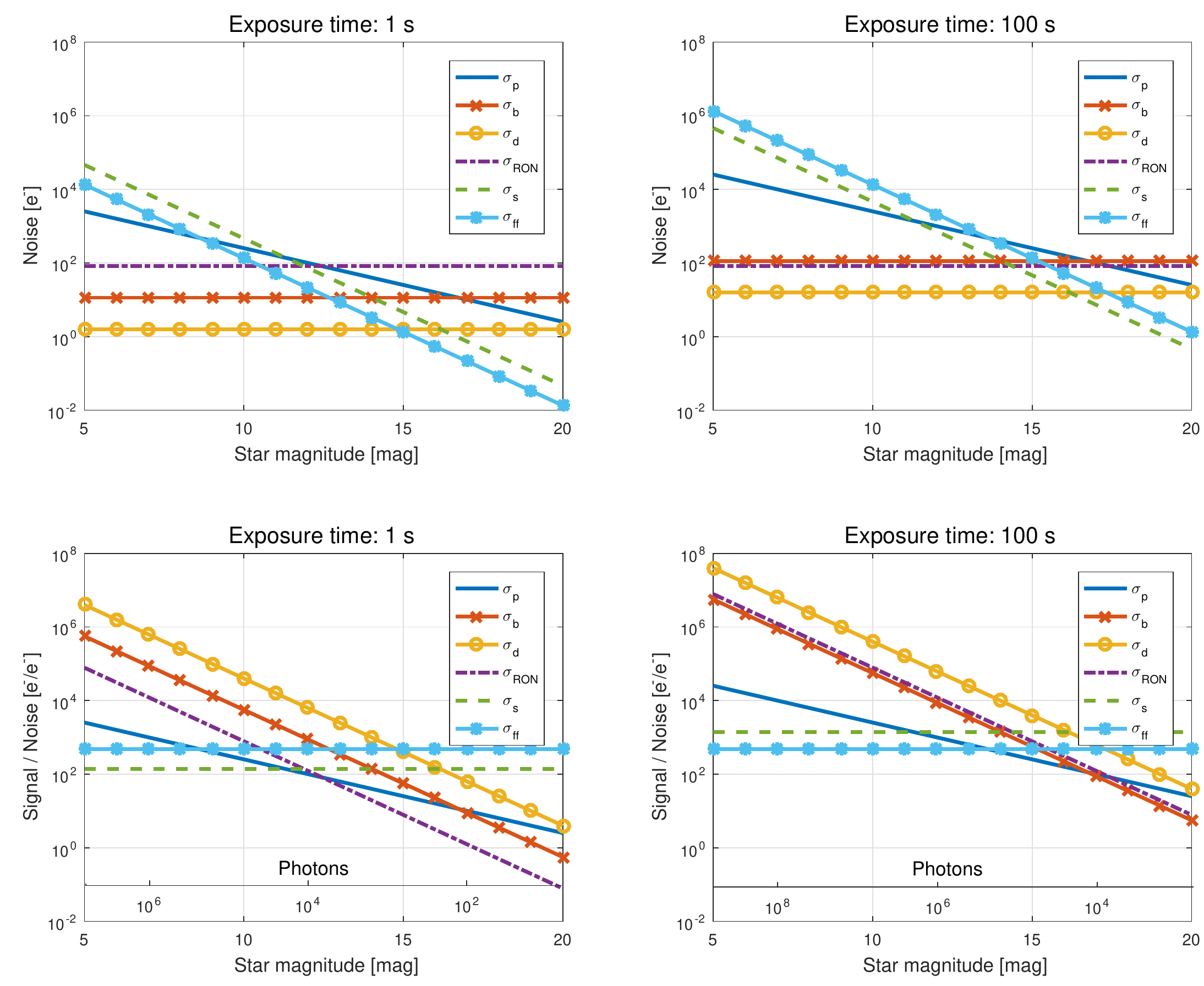}
	\caption{Theoretical dependencies of absolute (upper row) and relative (lower row) magnitude of various noise sources for a range of star magnitudes in V filter, for 1\,s and 100\,s exposures obtained with 0.5\,m telescope. Additional axis in the lower plots show the corresponding number of photons received from the object. (Color figures are available online).}
		\label{noiseEx}
\end{figure}

In Fig.~\ref{noiseEx} we present the absolute and relative magnitude for all the noise terms, for simulated 1\,s and 100\,s exposures with 0.5\,m telescope\footnote{The methodology of predicting the number of photons received from a star of given magnitude (per squared centimeter of aperture and per second) is provided by Cornell University Institute of Astronomy at \url{http://topics.sirtf.com/Astro4410/EstimatingPhotons} and was based on the following works: \cite{absphot1,absphot2,absphot3}.}. We assumed the following: observations in V filter (90\,nm width), exactly in the zenith $\gamma = 0$, readout noise $\sigma_{r} = 3$\,e$^-$, dark current $dD=0.005$\,e$^-$/pix/s, flat fielding residual noise $r_1=0, r_2=1$ \%, moderate seeing conditions FWHM = 2\arcsec, pixel scale 0.2\arcsec/pix and the scintillation coefficient $C_Y=1.5$. For the given FWHM, the photometric aperture covered $A_{pix}=510$ pixels. The value of the readout noise and the low intensity of dark current assumed in the simulations are achievable by nowadays state-of-the-art CCD cameras. 

In Fig.~\ref{noiseEx} it can be observed, that there are three types of noise sources: (1) independent of the object intensity ($\sigma_{RON}$, $\sigma_{d}$ and $\sigma_{b}$), linearly dependent ($\sigma_{RR}$ and $\sigma_s$) and the photon noise of counts from a source, $\sigma_p$, which shows a square root dependency. Of all the kinds of noise, only the readout noise is independent of exposure time. 

\begin{figure}
	\centering
	\includegraphics[width=\textwidth]{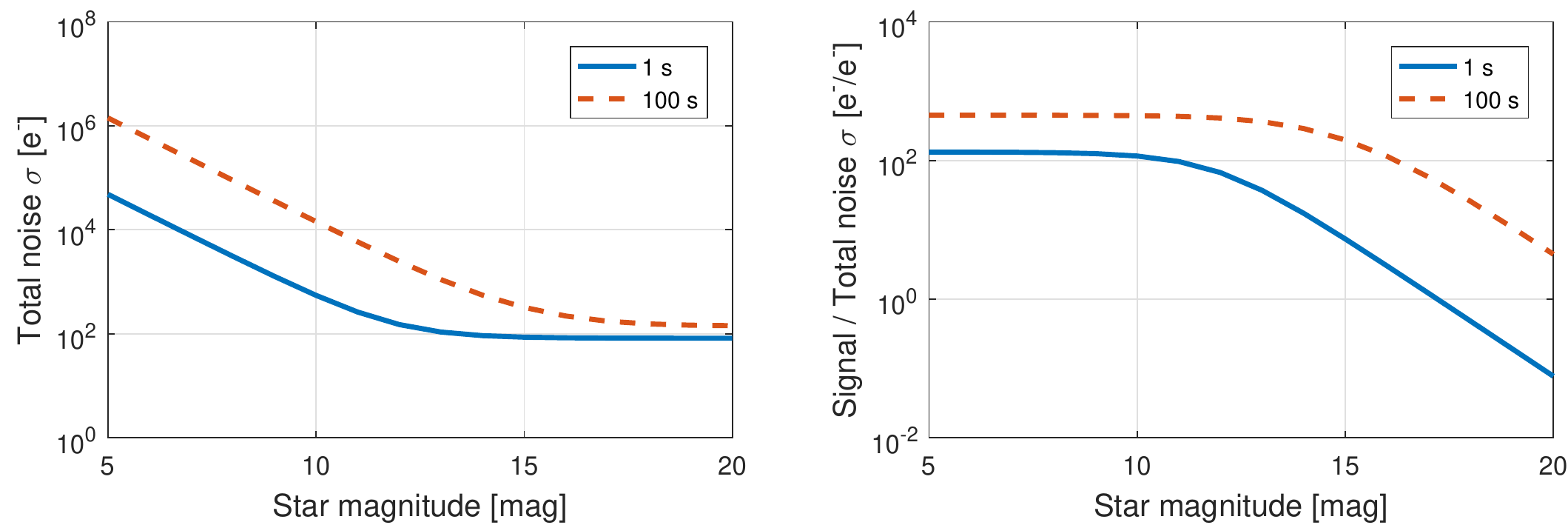}
	\caption{Dependencies of total noise $\sigma$ on stellar magnitude for 1\,s and 100\,s exposure with 0.5\,m telescope. The plots show the result of summation of noise sources presented separately in Fig. \ref{noiseEx}. (Color figures are available online).}
		\label{noiseEx2}
\end{figure}

The characteristics of corresponding total noise, $\sigma$, calculated for the noise sources presented separately in Fig.~\ref{noiseEx}, are showed in Fig.~\ref{noiseEx2}. Here, it can be seen that the flattening of error curves appears in both plots. For the absolute value of total noise (left plot), the flattening visible for dim objects appears due to the impact of the noises independent of the object intensity ($\sigma_{RON}$, $\sigma_{d}$ and $\sigma_{b}$). The flattening on the right plot limits the possible photometric accuracy of the brightest stars because of the presence of both flat-field ($\sigma_{RR}$) and scintillation ($\sigma_s$) effects.

\vspace{2mm}
\section{Effects of exposure time division}
\label{sec:efects}
In this section we concentrate on the impact of diversification of exposure times into a series of shorter sub-exposures. Lets assume that the time resolution of photometry is well determined and the exposure time available for a single measurement is assigned by $t_{total}$. Let us also assume that one knows the characteristics of the instrumentation (readout time, readout noise, dark current, quality of flat-fields, etc.) and the expected scintillation coefficient. For the sake of this study, we use the aperture photometry, in which the aperture covers $A_{pix}$ pixels including whole stellar profile (i.e. all the photons coming from the star are counted in the aperture). The signal to noise ratio (SNR) for a single measurement obtained in $t_i$ exposure time, is given by the following formula:
\begin{equation}
	SNR = \frac{S}{\sigma} =  \frac{S}{\sqrt{ \sigma^2_p + \sigma^2_b + \sigma^2_d  + \sigma^2_s + \sigma^2_{ff} + \sigma^2_{RON}}}.
\end{equation}
However, if the exposure time is divided (due to the saturation of the brightest stars) into $N$ shorter sub-exposures, the SNR of photometric outcome changes as follows:
\begin{equation}
\footnotesize{
	SNR_{s} =\frac{dS\sum_{i=1}^Nt_i}{\sqrt{(dS+dB~A_{pix}+dD~A_{pix}+dS^2\sigma^2_s \prime)\sum_{i=1}^Nt_i  + \sigma_{ff}^2 + N\sigma^2_{RON}}}.
	}
\end{equation}
Obviously, the total time available for obtaining single measurement cannot be exceeded:
\begin{equation}
	\label{timeeq}
	\sum_{i=1}^N (t_i+t_{R}) \leq t_{total}.
\end{equation}

Unfortunately, due to the division, the effective exposure time, in which the detector collects photons from the source, is reduced not by a single readout time, as it is for one long exposure, but by $Nt_R$. Moreover, the readout noise from pixels in the aperture is introduced $N$ times ($N\sigma_{RON}^2$). While for the brightest stars this has nearly negligible impact, for the dimmest ones it leads to significantly degraded photometric accuracy.

The only way to improve the photometric performance is to reduce the time spent on the readout and reduce the number of sub-exposure. One can try to reduce the readout time by increasing clocking frequency of image sensor, however it is always accompanied by increasing the readout noise due to the shorter time available for correlated double sampling \citep{P2012}. Our proposition is to divide the total time into less exposures at the expense of saturation of the brighter stars in some frames. Fortunately, these stars can be usually measured with a reasonable precision in shorter exposure/exposures.

Eventually, we introduce the definition the SNR for a single object if the total exposure time is divided into $N$ sub-exposures and the saturation is allowed:
\begin{equation}
\footnotesize{
	SNR_{s} =\frac{dS\sum_{i=1}^N\omega_it_i}{\sqrt{(dS+dB~A_{pix}+dD~A_{pix}+dS^2\sigma^2_s \prime)\sum_{i=1}^N\omega_it_i  + dS^2\big(\sum_{i=1}^N\omega_it_i\big)^2~r_{ff}^2 + \sigma^2_{RON}\sum_{i=1}^N\omega_i}},
	}
	\label{mainEQ}
\end{equation}
where $\omega_i$ represents the binary weight, which can be either 1 (no saturation) or 0 (saturation), depending on the presence of saturation of any pixel of a given object in $i$-th exposure. Obviously, the requirement presented in eq. \ref{timeeq} has to be met as well. Using the previous notation of a normalized profile $I_n$, the weight $\omega_i$ can be written as follows:
\begin{equation}
	\label{q_sat_eq}
	\omega_i = \bigg(dS~t_i~\max_{x,y} {I(x,y)} < Q_{sat}\bigg).
\end{equation}

To simulate exemplary photometric error curves, we adopted the characteristics of instrument and the observing conditions, as it was assumed in the previous section, providing additional information about the readout time $t_R$ = 10\,s and the saturation limit $Q_{sat}$ = 100\,000\,e$^-$. We also assumed that the stars of magnitude 11$\sim$18 are observed. From our model we determined that the exposure time, for which the brightest stars (11 mag) are nearly at the saturation level, is approximately 10\,s. We allowed the total exposure time of 100\,s, which resulted in a division into 5$\times$10\,s sub-exposures, while the remaining 50\,s had to be spent on the readout. The photometric accuracy for a given $s$-th star was computed according to the following formula:
\begin{equation}
	\Delta_{mag}(s) = 2.5\log{(1+1/SNR_s(s))}.
\end{equation}

\begin{figure}
	\centering
	\includegraphics[width=\textwidth]{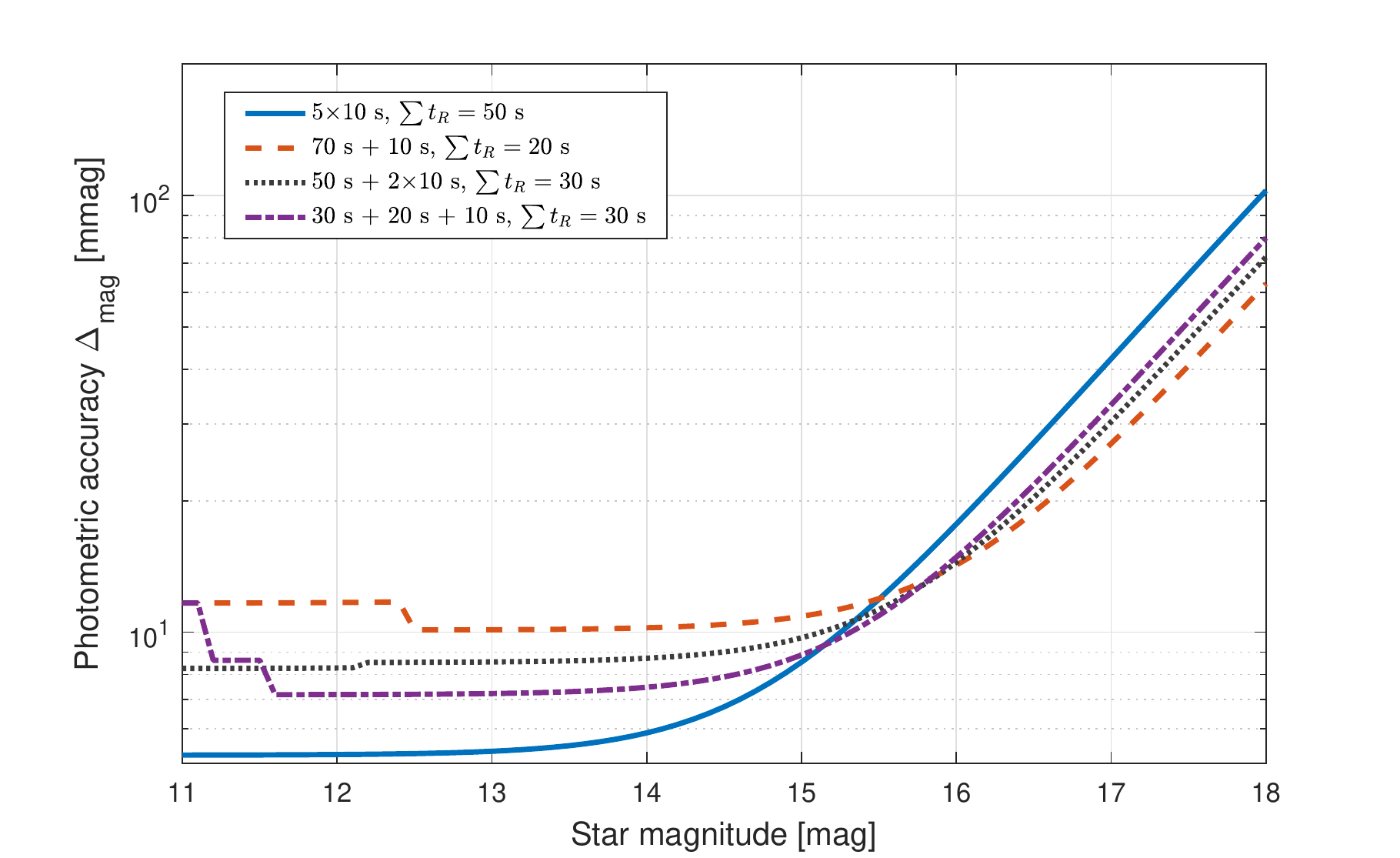}
	\caption{Comparison of photometric quality for various division of total exposure time. Only the solid line show the dependency for observations without saturation. (Color figures are available online).}
		\label{ncomp2}
\end{figure}

The results for four scenarios of exposure time diversification are presented in Fig.~\ref{ncomp2}. Several discontinuities visible in the dashed, dash-dotted and dotted lines, appear due to the saturation of a part of brightest objects in some exposures. These examples shows that the error curve for a range of magnitudes can be shaped by adjusting the number and the times of sub-exposures.

\section{Genetic algorithm}
\label{sec:GA}
As the SNR was defined for a single source, one has to decide how the photometric error curve should look like. From the perspective of photometric measurements, it seems to be reasonable to consider two possible goals: ($G_1$) the minimization of mean photometric error over whole stars set or ($G_2$) the maximization of the number of stars which achieve photometric precision better then predefined threshold value ($\Delta_{thr}$). While the first adjective would be useful for general-purpose photometry, the second proposition is dedicated for surveys aiming at detection of variability at magnitudes lower then specified threshold. These two targets  can be achieve by adjusting the number of sub-exposures $N$ and the vector of sub-exposure times $\{t_1,t_2,...,t_N\}$, so that the following is achieved:
\begin{equation}
	\label{optproblem1}
	G_1 = \sum_{s=1}^{M_{stars}}\Delta_{mag}(s) \xrightarrow[\{t_1,t_2,...,t_N\}]{N} \textrm{min}
\end{equation}
or
\begin{equation}
	\label{optproblem2}
	G_2 = \sum_{s=1}^{M_{stars}}(\Delta_{mag}(s)>\Delta_{thr}) \xrightarrow[\{t_1,t_2,...,t_N\}]{N} \textrm{max},
\end{equation}
where $M_{stars}$ is the number of stars measured in the field of view.

Prior to the optimization, it is required to analyze the field which will be observed. For this purpose, one can capture several frames, calibrate them and calculate the robust average, rejecting possible artifacts like the cosmic ray events. In the next step a standard photometric procedure, e.g. \href{http://www.star.bris.ac.uk/~mbt/daophot/}{DAOphot} \citep{DAOPHOT}, should be applied to locate the stars and define their estimated intensity. It is crucial to estimate the stellar profile for each detected star, so that it will be possible to predict possible saturation for a given exposure time according to eq.~\ref{q_sat_eq}. As a final outcome of such pre-observations, one should know the following set of information: (1) $dS$, $dB$ and $dS_{max}$ for each star, and (2) expected number of aperture's pixels, $A_{pix}$, calculated using actual width of stellar profile, FWHM. Additionally, the time available for registering single photometric point, $t_{total}$, has to be defined. In a well-calibrated telescope system, with characterized transfer function, known efficiency, and by obtaining actual seeing conditions, these all data can be estimated without performing pre-observations. 

The optimization proposed in this paper is performed in strongly nonlinear hyperspace, where small changes of inputs (sub-exposure times) may result in significant variability of the output ($G_1$ or $G_2$), mainly due to the discrete character of a binary weight $\omega$. Therefore, to obtain the optimal times of sub-exposures, we employed the genetic algorithm (GA), in which a set of solutions is gradually modified to reduce the defined error (eq.~\ref{optproblem1} or \ref{optproblem2}). This type of algorithms was proven to be capable of finding very good solutions for highly nonlinear problems \citep{GA1, GA2, GA3} as it is resistant to settling in local minimum. Additionally, we were encouraged by our recent successful application of GA-based approach for the observational purposes \citep{GAseeing}.

In the proposed GA, a single solution, so-called individual, is a set of $N$ sub-exposure times. We start the algorithm independently for defined $N$ value, thus it concentrates only on proper splitting the total exposure time into $N$ sub-exposures. Therefore, for each $N$ we obtain the best solution, which are eventually compared with each other to select the best times vector. The algorithm is proceeded according to the following steps:
\begin{enumerate}
	\item{$P$\,=\,1000 randomized individuals are created using the uniform distribution. Each individual is created by first drawing $N$ random values: $\{R_1,...,R_N\}$ and calculating the corresponding percentage of total exposure time: $\{R_1 / \sum{R},~...,~R_N / \sum{R}\}$ to compose the vector. Of course, the total exposure time has to be initially reduced by $N\cdot t_R$ to secure the time for readouts.}
	\item{The corresponding SNR is calculated for all stars and eventually the quality (eq.~\ref{optproblem1} or \ref{optproblem2}) is obtained for each individual. This process requires high computational time, especially if hundreds of sources are analyzed, therefore we perform it in parallel on all available CPU threads.}
	\item{$E$\,=\,50 best solutions (so-called elite) are found in the population $P$. The elite will be moved directly to the new population in next iteration.}
	\item{$M$\,=\,250 individuals are created by slight modification of the elite (mutation), which also survive to the next population. From each individual from the elite, 5 mutated versions are created by modifying the sub-exposure times with a vector of Gaussian random values. We used 1 second standard deviation, which allowed for achieving satisfactory results.}
	\item{The elite and its mutated versions are assigned to the next population, while the remaining 700 individuals are created fully randomly. The iterations are repeated defined number of times ($I$).}
 \end{enumerate}

Thanks to a significant number of random solutions created additionally in each iteration, the chance to find global optimum and leaving a local one is increased. The upper bound for $N_{max}$ can be easily determined by specifying the number of exposures required if none of the images is saturated. Increasing the number of sub-exposures above $N_{max}$ makes no sense, since the exposure time will be reduced and the number of readouts will be larger leading always to less collected charge and causing higher level of readout noise.

In the first step, the GA is run for all possible $N$ values using only $I$\,=\,10 iterations. As the initial guess of the optimal $N$ is found, the GA  is repeated only for this particular $N$ using larger number of iterations, $I$\,=\,100, to try to further improve the outcome. However, we observed that the improvement is only slight and does not exceed 5\% reduction of a goal function $G$.

\vspace{2mm}
\section{Experiment}
\label{sec:Experiment}
\subsection{Observations}
We tested our idea by comparing the photometric precision achieved by (1) dividing the time into non-saturated exposures and (2) by using the optimized ones. For this purpose we used two instrumental setups: 12'' Ritchey-Chretien telescope and 80\,mm apochromatic refractor, both mounted on \href{http://www.skywatcher.com/product/eq8-synscan/}{EQ8} paralactic mount. To have reasonably wide field of view, we attached \href{http://www.atik-cameras.com/product/atik-11000/}{ATIK 11000M} monochromatic, full frame cooled camera, which is dedicated for astronomical purposes. Its 11\,Mpixel sensor, \href{http://www.onsemi.com/pub/Collateral/KAI-11002-D.PDF}{KAI-11002}, achieves 13\,e$^-$ readout noise for a full-frame readout time of 30 seconds. The image sensor covered a field of  1$\degr\times$0.6$\degr$ (0.76\arcsec/pix) and 5$\degr\times$3.5$\degr$ (4.64\arcsec/pix), respectively for 12'' RC and for 80\,mm refractor. The thermal stabilization of the camera allows for calibrating the outcomes using averaged master dark frames. The observations were performed during eight nights in March/April 2017 at private observatory\footnote{observatory details: Kotulin, Poland: \href{http://wikimapia.org/\#lang=pl\&lat=50.450000\&lon=18.416667\&z=13\&m=b\&search=50\%20\%E2\%97\%A6\%2027\%200\%2057\%2000\%20N\%2018\%20\%E2\%97\%A6\%2025\%200\%2002\%200\%20E}{$50^\circ 27' 57''$N $18^\circ 25' 02'$E}.} located at relatively dark site about 30\,km from the \href{https://www.polsl.pl/en/Pages/Welcome.aspx}{Silesian University of Technology}. 

The observations were divided equally between two setups -- 4 nights with RC 12'' and 4 nights with 80\,mm refactor. In all observing runs, each 6-hours long, we used the following strategy. Initially, the mount was pointed toward a sky patch slightly on the west side of meridian to prevent the equatorial mount from flipping during the night. As the pre-observations we registered 10$\times$ 20\,s exposures which were calibrated using averaged flat-field and dark-frame and stacked employing sigma-clipping to reject cosmic ray events. On such calibrated images we performed source detection using DAOphot \citep{DAOPHOT} routines (detection with $\sigma$\,=\,9). All the saturated stars were rejected. Using the star images across the whole field of view, we utilized the PCA-based technique presented in \citep{PCA1,PCA2} to obtain the shape of stellar profile as a function of position on CCD. The normalized stellar profiles, $I_n$, were further utilized in the GA for calculating binary weights $\omega$.

Using the results of the analysis of pre-observations, we proceeded with GA optimization of exposure time division. We assumed the required time resolution of photometry as 300\,s which resulted in a series of $6\times20$\,s exposures in non-saturated mode (+$6\times30$\,s total readout time). The same target time-resolution was used for all the observations. As the goal function we used the mean photometric error, $G_1$~(\ref{optproblem1}). For simplicity, we assumed moderate scintillation conditions $C_Y$\,=\,2 which, as it was later determined, appeared to be close to the correct value. Since the telescope tracking worked very well ($<0.6\arcsec$ RMS error) and there was no field shifting, we assumed $\sigma^2_{ff1}=\sigma^2_{ff2}=0$. We used the circular aperture of radius three times the standard deviation of previously fitted Gaussian profile to cover whole stellar profile in the aperture. The intensities ($dS$) and the corresponding background levels ($dB$) for each detected star were determined by processing the images from pre-observations. The average rate of the dark current at the operating CCD temperature $-20^{\circ}$C was $dD=0.01$ $e^-/ \textrm{pix}/s$. The readout noise equaled to $\sigma_{RON} = 13$ e$^-$.

The \textit{seeing} in each night was determined by fitting the sharpest star images, in the central part of field of view, with Gaussian profile. The conditions were typical for the observing site and ranged between 1.5\arcsec \,$\sim$\,3\arcsec~FWHM. Obviously, the estimates of FWHM using images registered by the refactor were less precise due to the need of sub-pixel fitting. Additional information about the fields and the exposure times suggested by GA algorithm are given in Tab \ref{tableFields1}. 

The genetic algorithm indicated an optimized series of 3 exposures in each field and in both setups. Also, the times were not so different which implies that, despite the selection of fields with significantly different density of objects (see varying number of detected stars in third column of Tab. \ref{tableFields1}), the distribution of magnitudes was very alike. As it was expected, in each optimized series the GA decided to set the shortest exposure time to 20\,s to obtain the photometric data for the brightest star/stars in a field.

\begin{table}
	\centering
	\caption{Main information about the fields observed in the experiment.}
	\label{tableFields1}
	\begin{tabular}{ccccc}
		Field No. & Setup  &Stars detected  &Seeing (FWHM)  & Optimized series [s]  \\ \hline
		1 &RC 12''  &395  &2.02\arcsec  & 20\,/\,45\,/\,145\\
		2 &f = 2430\,mm  &130  &2.49\arcsec  & 20\,/\,31\,/\,159\\
		3 &1$\degr\times$0.6$\degr$ (0.76\arcsec/pix)  &370  &1.98\arcsec  &  20\,/\,43\,/\,147\\
		4 &  & 884 &3.11\arcsec  & 20\,/\,40\,/\,150\\
		\hline
		5 & refractor 80\,mm &  423 & 2.15\arcsec  & 20\,/\,32\,/\,158\\
		6 &  f = 600\,mm &   217 & 1.65\arcsec  & 20\,/\,52\,/\,138\\
		7 & 5$\degr\times$3.5$\degr$ (3.09\arcsec/pix) &  549 & 1.92\arcsec  & 20\,/\,60\,/\,130\\
		8 &  & 329 & 2.31\arcsec  & 20\,/\,34\,/\,156\\
		\hline
	\end{tabular}
\end{table}

\subsection{Data analysis}
To obtain the photometric results  we used the circular aperture photometry with a constant radius. As it was stated before, the radius was a tripled standard deviation of average fitted Gaussian profile. Since the profiles in images registered by 80\,mm refractor were very narrow, we decide to use partial pixels in the aperture. This means that the signals from pixels falling partially into aperture were added proportionally to the area of this pixel falling into the aperture. For each star, the charge was summed over all images in 300\,s series creating a single photometric outcome. 

Since the brightest stars were registered only in the shortest exposures, while to dimmer ones could be measured in each image, we decided to work with fluxes expressed in e$^-$/s which were further multiplied by 300\,s to estimate final charge. Due to the presence of many variable stars in experimental fields, which showed gradual decrease or increase of intensity over observed period, we decided to estimate the photometric noise as a spread of lightcurve points around the linear fit. This allowed for including much more stars in our comparison. It was especially important for field~4, which included the open cluster M67, wherein most of the brightest stars are variable. Eventually, the noise level in each lightcurve was compared with the theoretical values indicated by eq.~\ref{mainEQ}.

\begin{figure}
	\centering
	\includegraphics[width=\textwidth]{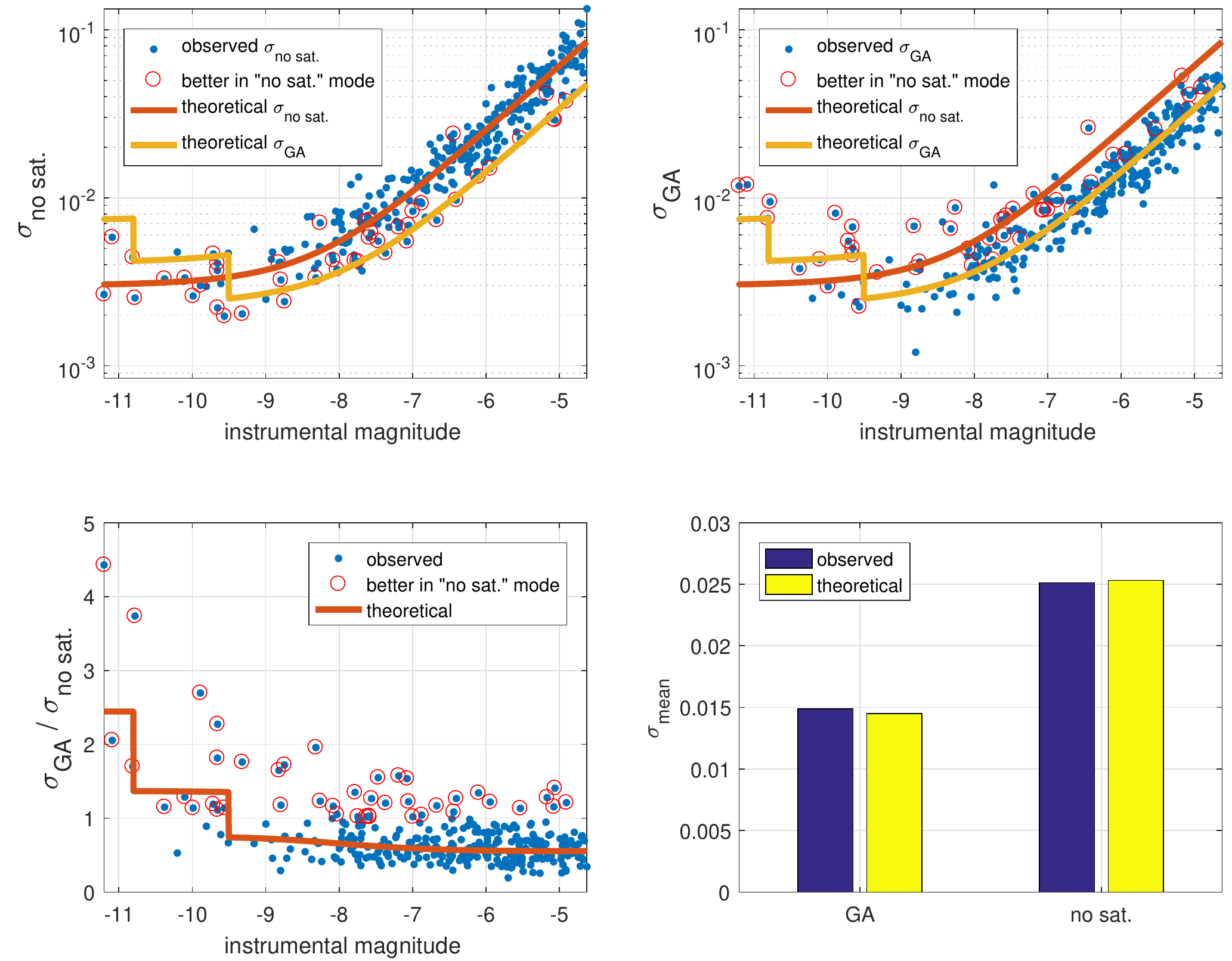}
	\caption{Comparison of photometric accuracy for standard (no saturation) and GA-based exposure time diversification in exemplary field (field 1). The upper plots show the photometric errors for both modes (standard -- left, GA-based -- right) compared with the theoretical expectations (solid lines). The stars which showed lower accuracy in GA mode, are annotated with red circles. The expected and obtained overall photometric performance expressed by the mean spread of photometric points, $\sigma_{mean}$, is given in the bar plot. (Color figures are available online).}
		\label{exampleField3}
\end{figure}

Exemplary detailed results from the analysis of field 1 are presented in Fig.~\ref{exampleField3}. Two upper plots show the photometric precision achieved by a standard (no saturation) and GA-optimized series, on left and right side, respectively. The theoretical error curves are indicated by thick red and orange lines. It is visible that the points follow the expected dependencies. Importantly, the assumed scintillation coefficient $C_Y$\,=\,2  appeared to fit well the noise floor visible in both error plots.

As it was expected, the photometric accuracy of the brighter stars decreased at the expense of more precise photometry of dimmer stars. By additional red circles, in Fig.~\ref{exampleField3}, we indicated the stars whose photometry degraded after applying \mbox{GA-optimized} exposures. They included nearly all of the brightest stars and, surprisingly, a small fraction of dimmer stars. They, however, turned out to be the outliers as they showed lower then expected noise in the standard mode while exposing enormously high one in the GA-based observations (these stars are indicated by red circles on all three plots). The comparison of the overall photometric quality expressed by $\sigma_{mean}$ is presented in bar plot in Fig. \ref{exampleField3}. The mean errors proved very good agreement between the experimental outcomes and the theory. The superiority of GA optimized exposures in this field was evident ($\sigma_{GA}=0.015$\,mag, $\sigma_{no. sat.}=0.025$\,mag).

\begin{figure}
	\centering
	\includegraphics[width=0.49\textwidth]{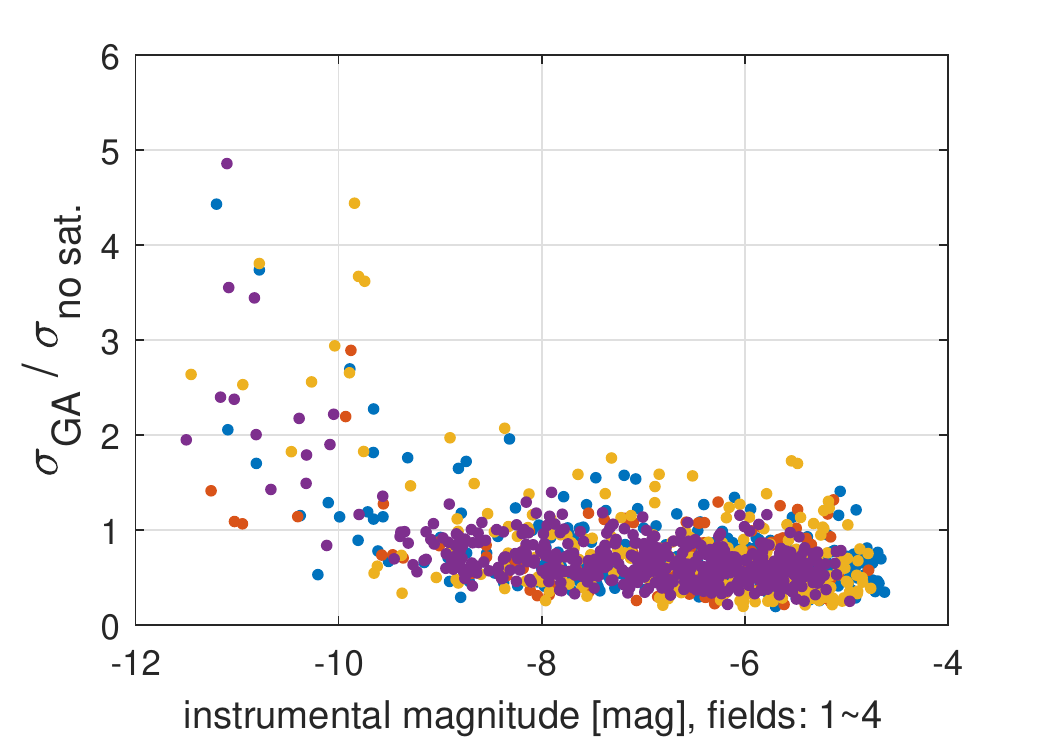}
	\includegraphics[width=0.49\textwidth]{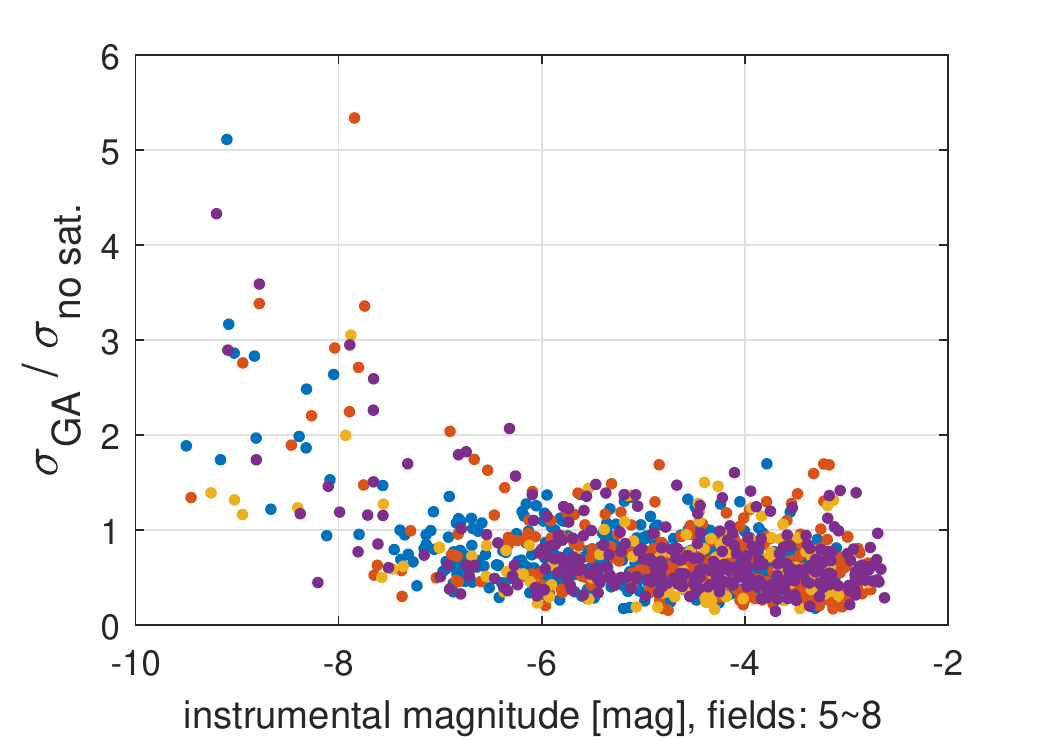}
	\caption{Relation between photometric accuracy for standard and GA-based observing modes ($\sigma_{GA}/\sigma_{no~sat.}$) for all 8 fields observed in the experiment. The results are divided into two plots according to used instrumental setup: left -- RC 12'', fields 1$\sim$4; right -- refractor 80\,mm, fields 5$\sim$8. (Color figures are available online).}
		\label{all_comp1}
\end{figure}

A comprehensive view on the spread of photometric quality for all fields and both observing modes is presented in Fig.~\ref{all_comp1}. The results are divided into two plots since the instrumental magnitudes were different in both observational setups due to the much narrower profiles in 80\,mm refractor. The four employed colors in both plots represent different fields. It is visible that after applying GA-based exposures the noise of dimmer stars (i.e. $>-9$ mag for left plot and $>-7$ mag for the right one) is significantly reduced while the brightest stars have the noise elevated. The same desirable relation is visible in both observational setups.

\begin{figure}
	\centering
	\includegraphics[width=1\textwidth]{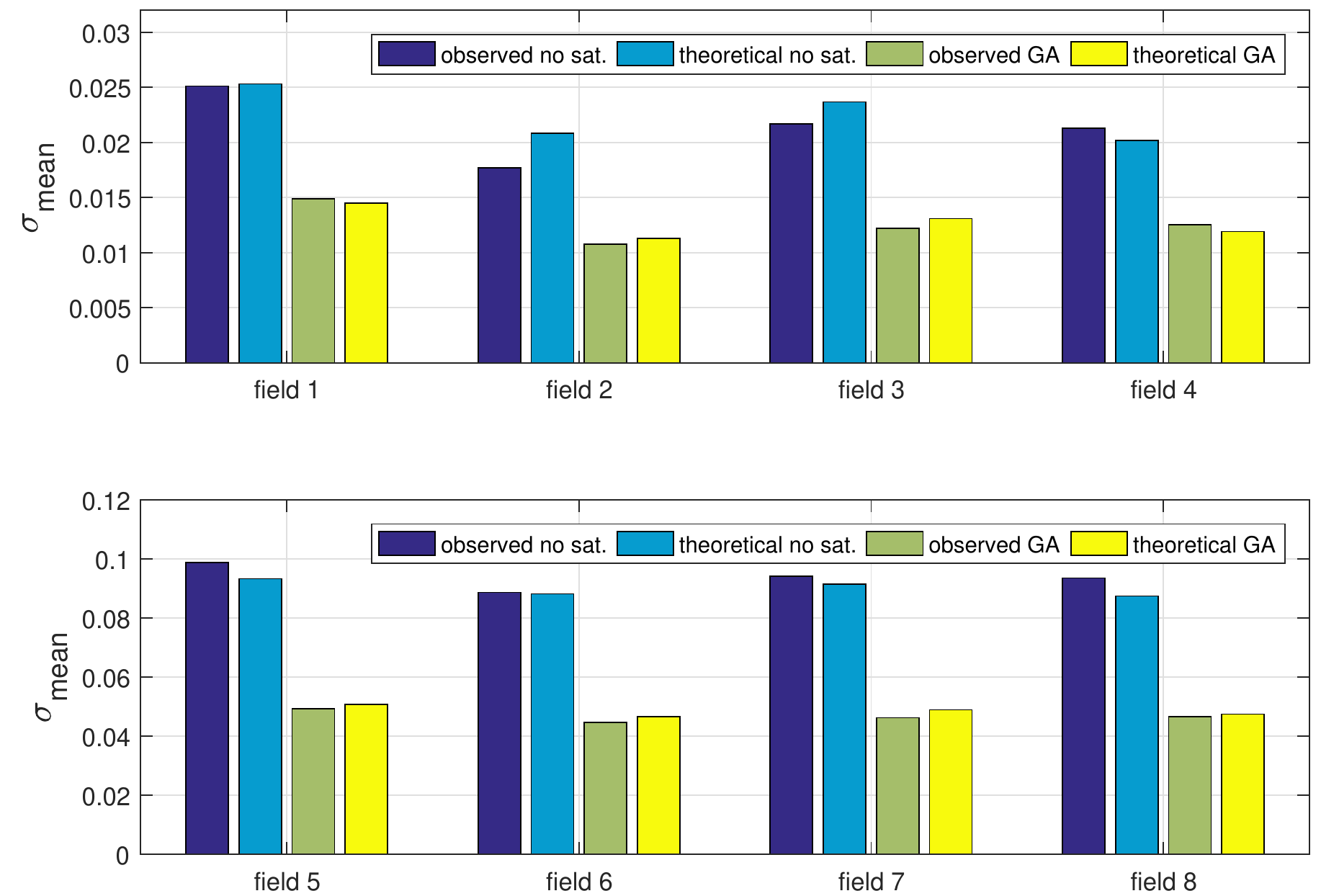}
	\caption{Comparison of expected and observed mean photometric error, $\sigma_{mean}$, for all 8 experimental fields. The upper plot corresponds to the data from RC 12'' telescope, while the lower one shows results obtained by the 80\,mm refractor. (Color figures are available online).}
		\label{bar_plots_all}
\end{figure}

The bar plots in Fig.~\ref{bar_plots_all} compare the theoretical and actual mean precision, $\sigma_{mean}$, calculated over all stars in each field. The theoretical values agree well with the observed outcomes. In both setups we noticed approximately twice lower mean spread of data points when using GA-based exposures. 

\section{Conclusions}
\label{sec:Conclusions}
Wide field photometry is the everyday routine for plenty of survey telescopes, all over the word. Unfortunately, when registering images with a constant exposure time one optimizes the photometric quality only of the brightest objects. In some cases, all the stars in the field, both bright and dim, are of equal interest of the astronomers.

To better balance the photometric precision between all the stars in the field, we proposed a novel approach of the selection of exposure times. Knowing elementary characteristics of equipped camera and assuming available time for measurements, it is possible to divide the total exposure time into several sub-exposures, in which the saturation of objects is allowable. By means of simple genetic algorithm, we were able find the division of time, which can optimize various photometric goal functions. It can be either the minimization of the mean photometric error or the maximization of the number of stars registered with an assumed precision. Thanks to our method, it was possible to improve the photometric quality of dimmer stars at the expense of lowering accuracy of the brightest ones.

In the paper we also described and characterized numerous sources of noise in photometric measurements with CCD image sensors. This noises should be carefully investigated in the employed instrumentation to maximize the efficiency of the proposeed GA-based technique.

The standard and GA-based observational strategies were tested in real observations of a eight random stellar field using two telescope setups. The photometry of registered images proved that the accuracy follows the predicted error curves. Importantly, by using GA-based exposures the mean photometric error was reduced to the level consistent with theoretical calculations. 

Our approach can be easily implemented and utilized in the operational pipelines of survey telescopes to enhance their photometric outcome and to improve the time distribution. The proposed strategy may be also of special interest of space survey missions, whose lifetime is usually limited by the degradation of CCD sensor due to the proton radiation. The reduction of the number of exposures is also valuable considering the power issues and data transfer limits of space telescopes.

\acknowledgments
\label{sec:ack}
{\bf Funding}.\quad Polish National Science Center grants no: 2012/07/B/ST9/04425, 2016/21/D/ST9/00656 and 2016/21/N/ST9/00375.

\bibliographystyle{aasjournal}
\bibliography{biblio}

\end{document}